\documentclass[12pt,preprint]{aastex}

\newcommand{\todash}{\,--\,}
\newcommand{\ic}{I$_{\rm c}$ }

\newcommand{\be}{\begin{equation}}
\newcommand{\ee}{\end{equation}}
\newcommand{\beq}{\begin{eqnarray}}
\newcommand{\eeq}{\end{eqnarray}}
\usepackage{amsmath}

\shorttitle{Local Frequency Shifts Between $V$ and $I$}
\shortauthors{Tripathy et al.}

\begin{document}


\title{Comparison of High-degree Solar Acoustic Frequencies and Asymmetry between Velocity and  
Intensity Data}

\author{S. C. Tripathy}
\affil{National Solar Observatory, Tucson, AZ 85719, USA}
\email{stripathy@nso.edu}

\author{H. M. Antia}
\affil{Tata Institute of Fundamental Research, Homi Bhabha Road, Mumbai 400 005, India}
\author{K. Jain and  F. Hill}
\affil{National Solar Observatory, Tucson, AZ 85719, USA}

\begin{abstract}

Using the local helioseismic technique of ring diagram  we analyze the frequencies of 
high--degree {\it f}- and {\it p}-modes derived from both velocity and continuum intensity data 
observed by MDI.
Fitting the spectra with asymmetric peak profiles, we find that the asymmetry associated with 
velocity line profiles is  negative for all frequency ranges agreeing with previous observations 
while the asymmetry 
of the intensity profiles shows a complex and frequency dependent behavior. 
We also observe systematic frequency differences between intensity and velocity 
spectra at the high end of the frequency range,  mostly above 4~mHz. 
We infer that this difference arises from the fitting of the 
intensity rather than the velocity spectra. 
We also show that the frequency differences between intensity and velocity  do not vary significantly 
from the disk center to the limb when the spectra are fitted with 
the asymmetric profile and conclude that only a part of the background is correlated 
with the intensity oscillations. 
\end{abstract}


\section{Introduction}

Different helioseismic instruments, both from ground and space, observe the Sun in different observables.
Due to the different techniques used by these instruments, it is possible to measure the solar oscillations
simultaneously either as variations in the photospheric velocity or as intensity fluctuations. 
It is therefore important to confirm that the oscillation mode parameters measured from both the intensity and 
velocity agree with each other to a high 
degree of precision. However, the initial measurement of low degree {\it p}-mode frequencies 
from Doppler velocity ($V$) and continuum intensity (I$_{\rm c}$) observations from Michelson Doppler Imager (MDI)
instrument on board {\it Solar and Heliospheric Observatory} (SOHO) showed 
systematic differences. A comparison of 108-day power spectra between $V$ and \ic
showed a weighted 
mean difference of $-0.1~\mu$Hz for $\ell=0$, and $-0.16~\mu$Hz for $\ell=1$ modes  \citep{toutain97}.  
Since the apparent frequency shift between an oscillation observed in velocity and intensity 
cannot be a property of the mode, it must arise from systematic errors
while calculating the frequencies from the observed power spectra. Hence it was argued 
that the source of the systematic difference could be due to the opposite asymmetry effect  
seen between the velocity and intensity power spectra \citep{duvall93}.   
\cite{app98} also presented a similar evidence using VIRGO and SOI/MDI data. 
Around the same time \cite{harvey98}  
reported that the intermediate degree modes observed in $V$ and total spectral intensity also show 
different central frequencies and observed that the apparent differences could be as large 
as 50~$\mu$Hz close to the acoustic cut-off frequency. However, the analysis of \cite{nigam98b},
using an asymmetric line profile-fitting formula, illustrated that the 
frequency difference between $V$ and \ic in the 
intermediate degree range is much smaller compared to that obtained by fitting a symmetric Lorentzian 
profile. Using the same asymmetric line profile-fitting formula, 
  \cite{toutain98} re-analyzed the data from MDI and showed that the  frequency differences 
between $V$ and \ic are considerably reduced.
   \cite{gav99} have also analyzed data from different 
   instruments and  have argued that the reported  frequency shift  
is merely an artifact of the reduction technique. 

Renewed interest in the topic began when local helioseismic techniques were developed  to study the properties of 
high-degree modes in localized regions.  \cite{bab01} analyzed azimuthally averaged (2-d) power spectra 
and inferred  that 
the eigenfrequencies obtained using the asymmetric peak profiles agree reasonably well with each other 
while the use of
symmetric profiles gives significant differences between frequencies computed using
continuum intensity and velocity or continuum intensity and line-depth spectra.  
In order to gain further information for high-degree and high-frequency modes, \cite{jain03} analyzed 
the high-resolution GONG+ data.  These authors also compared the azimuthally averaged power spectra 
 of a region centered on the equator and reported 
that the frequency dependence of the frequency shift between $V$ and I is negligible below the acoustic cutoff
frequency around 5.3 mHz and substantial above the cutoff. These results supported the finding of \cite{harvey98}.
However, the conclusion is based on the visual comparison of the peak frequency 
of the power spectra and  may not necessarily be a true measure of the shift due to 
the reversal of the asymmetry and different background power between $V$ and \ic spectra. 

It is now well established that line asymmetry of the solar power spectra alters the peak 
frequencies that are obtained under the assumption that the lines are symmetric (e.g. \cite{ab99, ba00}. 
However, the cause 
of the opposite asymmetry between the velocity and intensity spectra
still remains inconclusive.  The current understanding is that the reversal in
the sign of asymmetry between the $V$ and \ic spectra is due to the 
influence of the solar background noise that is correlated with the source \citep{rv97, nigamall98, severino01} and 
the level depends on the characteristic granulation. On the other hand,  
the numerical simulation \citep{geo03} indicates that the reversal is produced by the radiative 
transfer effects. Since the physics of the correlated noise is not yet fully understood and 
the spatial leakage 
signature for $V$ and I is  different due to their center-to-limb variations,  our objective is 
 to examine the frequency dependence of the observed asymmetry and 
differences in eigenfrequencies between velocity and intensity observations   
as a function of the radial distance from the disk center to the limb.
A preliminary investigation of a similar nature using azimuthally averaged power spectra
is reported in \cite{sct06}. However  the present  analysis differs from all earlier ones since here 
we use the three-dimensional (3-d) 
power spectrum, which is associated with flow fields, while the azimuthally averaged spectrum 
has no flow fields associated with it. 

The rest of the paper is organized as follows: 
We briefly describe the data and analysis technique in Section 2, while the results 
are described in Section~3. Finally, we summarize the main conclusions in Section~4. 

\section{Data and the Technique}
We use the Dopplergrams and continuum intensity images obtained 
by the MDI instrument during the period of 1997 May 19 \todash\ 21 when solar 
activity was near minimum.   We have chosen 4 regions centered at heliographic 
longitudes of 
$0\degr$, $15\degr$,  $30\degr$, and $45\degr$; all centered at the equator.  
The spatial extent of each of the localized region covers 
$256\times256$ pixels in heliographic longitude and 
latitude and gives a resolution of 0.01616 {Mm}$^{-1}$. 
 Each region is tracked for 4096 minutes, which gives a 
frequency resolution of 4.07~$\mu$Hz. The standard ring diagram technique \citep{hill88} was used to 
obtain the power as a function of ($k_x$, $k_y$, $\nu$). 

To extract the frequencies and other mode parameters, the three-dimensional power spectrum is fitted with 
a model with asymmetric peak profiles  of the form
\begin{eqnarray}
&&P (k_x, k_y, \nu) = \frac{e^{B_1}}{k^3} + \frac{e^{B_2}}{k^4} \nonumber \\ 
&+& \frac{exp(A_0 + ( k - k_0)A_1 + A_2(\frac{k_x}{k})^2 
+ A_3 \frac{k_xk_y}{k^2})S_x}{x^2 +1}
\end{eqnarray}
where 
\begin{eqnarray}
x &=& \frac{\nu - ck^p - U_xk_x -U_yk_y}{w_0 + w_1(k - k_0)}, \\
S_x &=& S^2 + (1 + Sx)^2.
\end{eqnarray}
Here $P$ is the oscillation power for a mode with a temporal frequency $\nu$ and
the total wavenumber $k=\sqrt{k_x^2+k_y^2}$. The peak power in the mode is represented 
as $\exp(A_0)$ while the  coefficients $A_1$ to $A_3$ account for the 
variation in power with $k$ and along the ring.  The terms involving $B_1$ and $B_2$ define the 
background power and  $w_0$ is the mode width.  
The term $ck^p$ gives the mean frequency, and this form is chosen as it gives satisfactory fits to the mean 
frequency over the whole fitting interval.  The parameters $S$ controls the asymmetry in the peaks, 
and the form of asymmetry is the same as that used by \cite{nigam98b} i.e. the parameter 
is positive for positive asymmetry and negative for negative asymmetry. By setting $S=0$ we can also fit 
symmetric Lorentzian profiles. 
  A more detailed description of the parameters and fitting
procedure is described  in \cite{bat99} and \cite{ba99}. The 13 parameters ($A_0$, $A_1$, $A_2$, $A_3$, $B_1$, $B_2$,  
$c$, $p$, $S$, $U_x$, $U_y$, $w_0$,  $w_1$)  
are determined by fitting the 
spectra using the maximum likelihood approach \citep{and90}. In this work we express $k$ in units of 
$R_{\sun}^{-1}$ so that $k$ is identified  with the degree $\ell$ of the spherical harmonic of the 
corresponding global mode. 

\section{Results and Discussions}
\begin{figure}           
\centering
\plotone{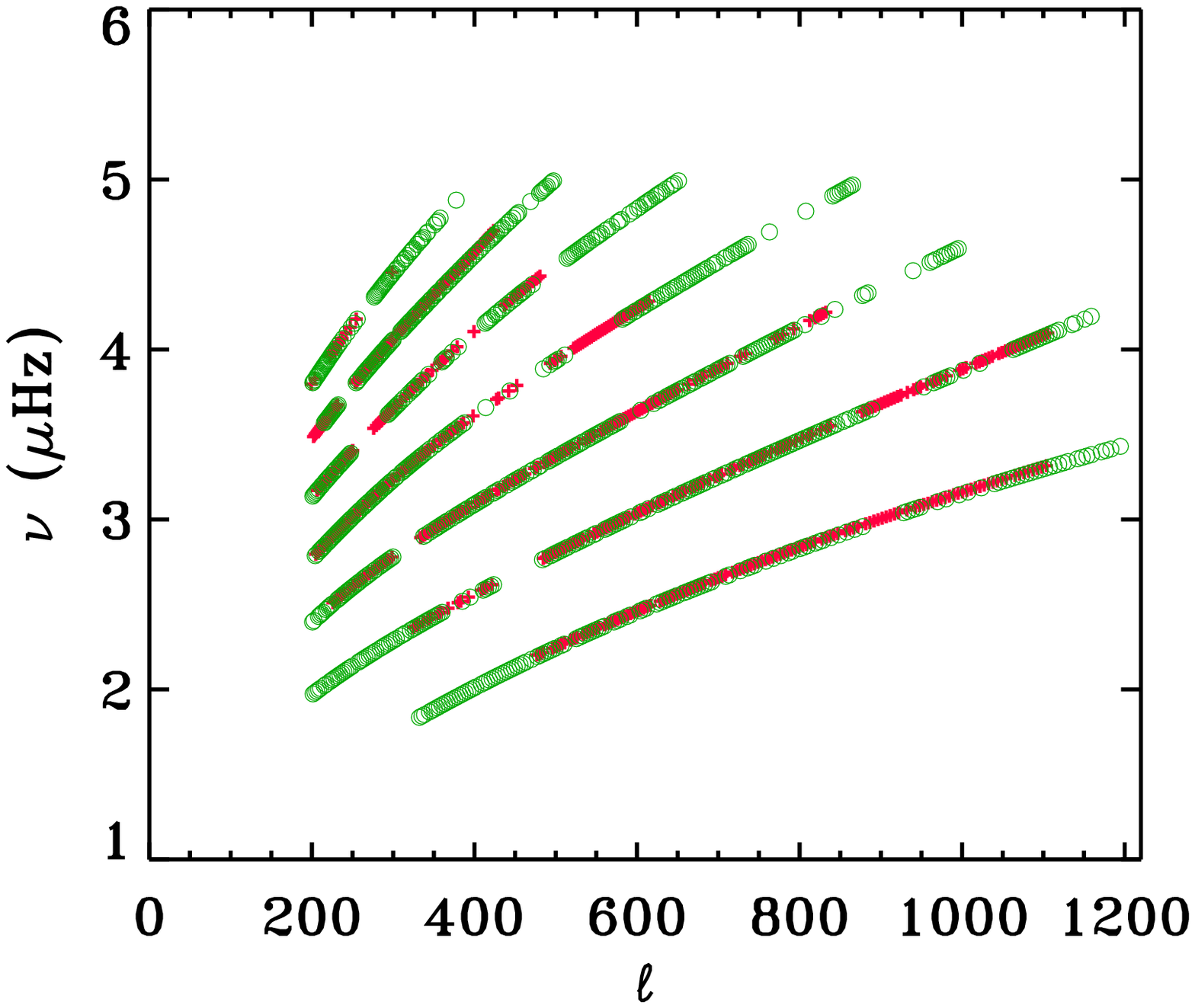}
\caption{The $\ell\, -\, \nu$ diagram constructed 
from the asymmetric fits to disk center spectra for velocity (circles) and continuum intensity (pluses).  
 \label{fig1}} 
\end{figure}

Figure~\ref{fig1} shows a
characteristic $\ell - \nu$ diagram obtained from the fits 
to the power spectra at the disk center 
using asymmetric profile corresponding to Eq.~(1). 
In general we obtain about 750  and 600 modes  
from the velocity  and intensity spectra respectively. Intensity spectra normally yields 
less number of modes due to the inherently lower signal-to-noise ratio. The number of fitted modes 
also decreases for regions that are away from the disk center. Comparing the number of modes 
between  symmetric and asymmetric fits we find that the symmetric fit yields more modes. This implies that 
the inclusion of
the asymmetry in the fitting formula reduces the number of modes that are successfully fitted.
This could be due to some cross-correlation between $S$ and other parameters of the model, particularly, 
the background.   Although, we fit the spectra up to radial order of $n = 6$, we restrict the analysis up 
to $n = 4$ modes since higher order radial modes have large estimated fitting errors.   

\subsection{Asymmetry of Peak Profiles} 
\begin{figure}           
\centering
\plotone{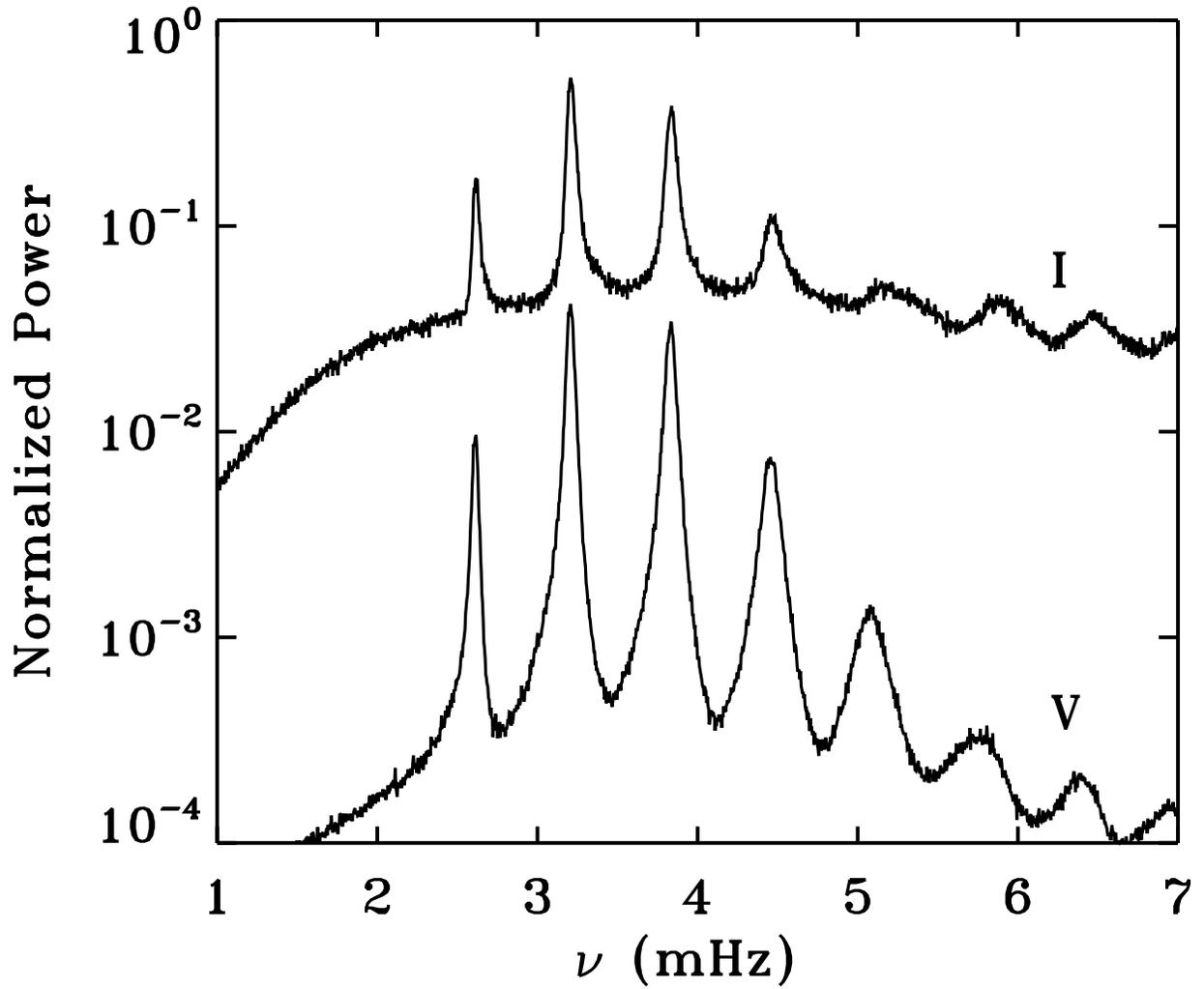}
\caption{Normalized power for $\ell$ = 675 corresponding to the azimuthally averaged and normalized 
velocity and continuum intensity disk center power spectra.  
 \label{fig2}} 
\end{figure}

Physically, asymmetry is a result 
of an interaction between an outward-directed wave from the source and a corresponding 
inward-directed wave that passes through the region of wave propagation \citep{duvall93}. 
To illustrate the peaks in the power spectra and to visualize the asymmetry associated 
with different observables, we show an example of the azimuthally averaged power 
 for $\ell=675$ corresponding to the disk center spectra for $V$ and \ic in Figure~\ref{fig2}. 
It is  apparent that the line profiles
are asymmetric. At low frequency, the asymmetry agrees with the known results; the velocity spectrum 
has negative asymmetry i.e. more power on the lower frequency side of the peak while 
the continuum intensity spectrum show positive asymmetry {i.e.} more power on the 
higher frequency side of the peak. This reversal of asymmetry is believed to arise 
from the addition of correlated noise to the amplitude spectra \citep{nigamall98}. 
 In addition, we notice that the asymmetry associated with velocity spectra  
appears to be reduced or even reversed at higher frequencies. 
Visual inspection of the azimuthally averaged power spectra show a comprehensible 
frequency shift of  the central frequency between $V$ and \ic  above the cutoff frequency of 5.3~mHz.  
We also observe  
that the velocity power is higher near the disk center and decreases toward the 
limb in agreement with the predominately radial nature of the oscillatory velocity field.
 
\begin{figure}           
\centering
\plotone{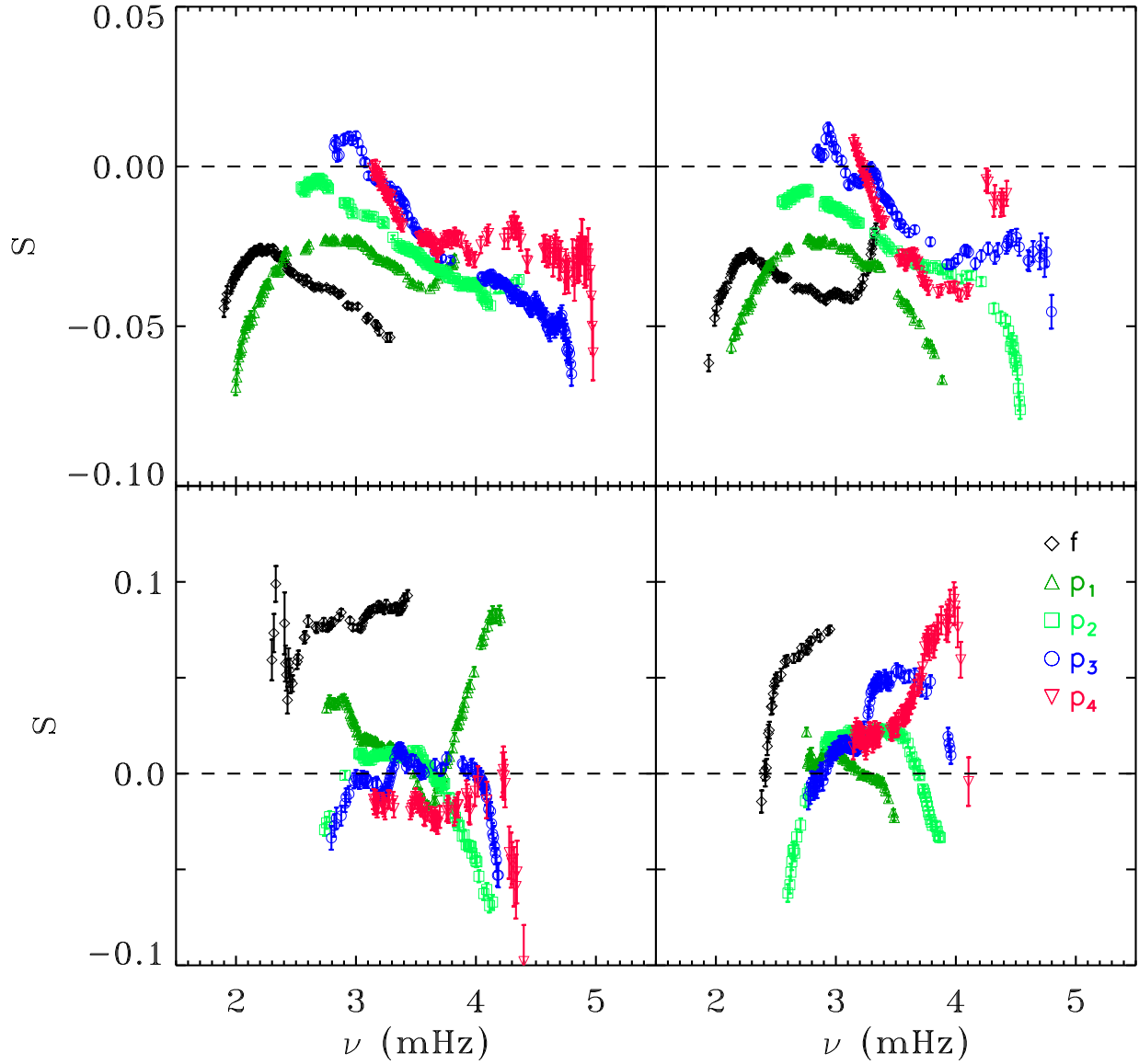}
\caption{Asymmetry parameter $S$ obtained from fits  to velocity (top panel) and continuum intensity   
(bottom panel) spectra at two different locations. 
The left and right panels refer to the regions at disk center  
 and at a longitude of 45$\degr$ E, respectively. 
The symbols represent different radial orders and are explained in the right bottom panel.   
 \label{fig3}} 
\end{figure}

Figure~\ref{fig3} shows the asymmetry of the line profiles as a function of frequency 
obtained  from  fitting  the three dimensional power spectra at disk center and 
45\degr\ E longitude. 
 As expected, this parameter is predominantly negative for all modes in the velocity spectra  (upper panels) 
indicating that there is more power on the low frequency side of the peak. The asymmetry is   
minimum around 3~mHz 
while at low end of the frequency range the asymmetry of {\it p}$_2$ and higher order radial modes   
 is more likely to be zero within the estimated errors.  Thus, in this frequency range the mode 
frequencies obtained from asymmetric and symmetric profiles should be nearly identical. 
Surprisingly, we do not observe the asymmetry to be reversed at higher frequencies as noted above and 
also in \cite{sct06}.
 
The asymmetry for the intensity spectra (lower panels of Figure~\ref{fig3}) appears to be more complex
and in general these are larger than the velocity asymmetry. For 
{\it f} and {\it p$_1$}-modes, this parameter is primarily positive (more power on the higher frequency 
end of the peak) and increases with frequency. For higher order {\it p}-modes, asymmetry 
 is negative both at low- and high-end of the frequency range 
and is likely to be insignificant or zero at other frequencies. Thus the asymmetry is different 
than what is observed for low-degree modes \citep{toutain98}, intermediate degree modes 
\citep{nigamall98}, and high-degree modes obtained from the azimuthally averaged spectra \citep{sct06}.  
Although most of these differences may be due to frequency, 
we do not rule out the possibility that the form of asymmetry used in the model are simplified 
and may not  represent the real profiles adequately at all frequencies.
The asymmetry may also weakly depend on $\ell$ \citep{ba99} and this dependence has been neglected 
in the fitting formula. Finally, we note that the asymmetry of both velocity and  intensity 
spectra do not show any significant changes at different locations on the solar disk. 

\subsection{Comparison of asymmetry between 2-d and 3-d fitting}
\begin{figure}           
\centering
\plotone{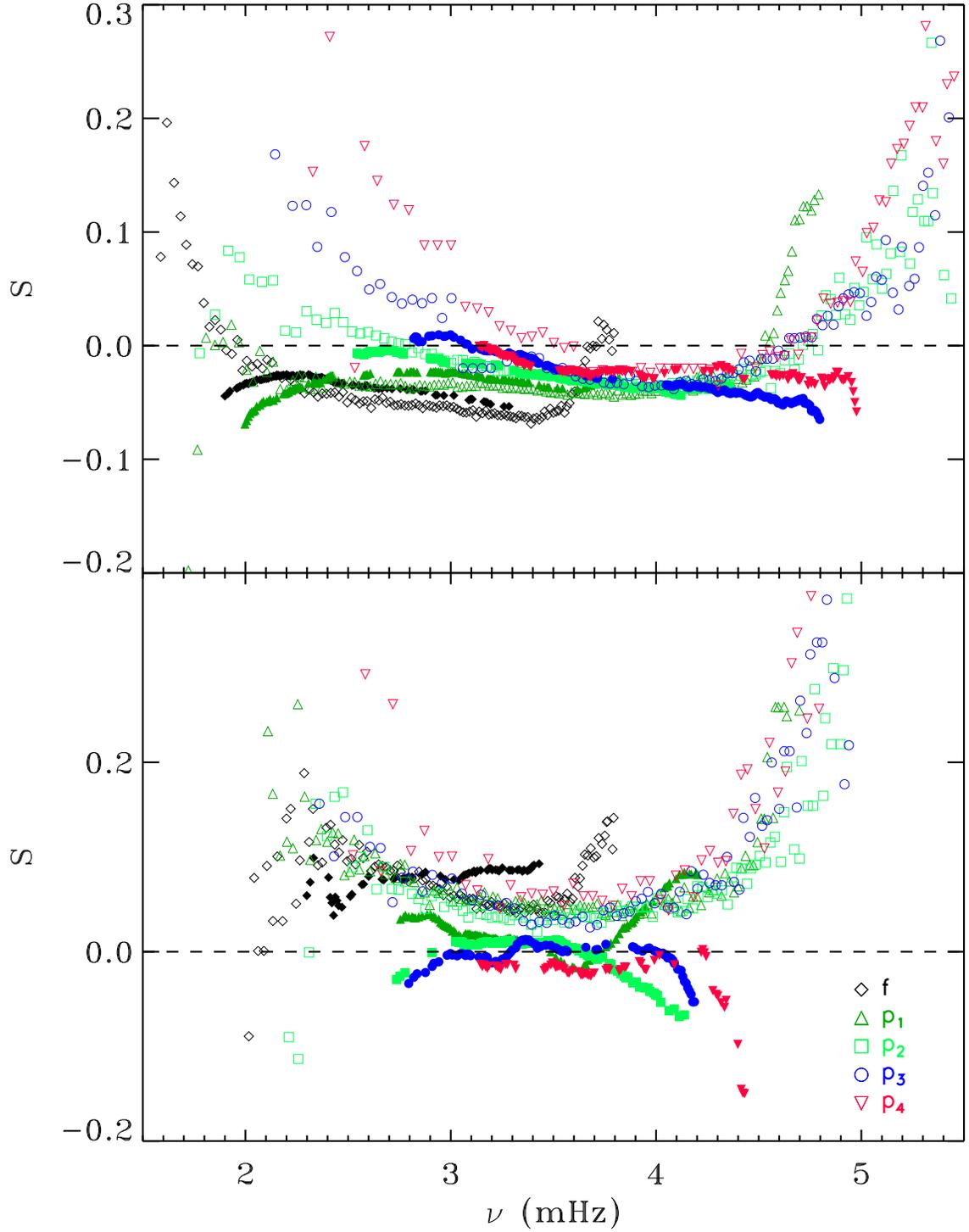}
\caption{Comparison of the asymmetry parameter obtained from fits  to three-dimensional 
(filled symbols) and azimuthally averaged spectra (open symbols) at disk center. The top panel 
is for velocity and bottom panel is for continuum intensity. The errors are not shown for clarity. 
\label{fig4}} 
\end{figure}

It is interesting 
to compare the asymmetries from 3-d spectra with those obtained from azimuthally averaged spectra. 
Figure~\ref{fig4} illustrates the asymmetries obtained from both the fits. 
While the asymmetry observed in intensity spectra
obtained from 2-d fits are entirely positive, the behavior is more complex in the 3-d spectra. On the 
other hand, asymmetry associated with 3-d velocity spectra is entirely negative while for 2-d spectra,
it reverses sign at high end of the frequency range.  Thus significant differences are 
seen between the asymmetry obtained from fitting the three-dimensional and azimuthally 
averaged power spectra. 

\begin{figure}           
\centering
\plotone{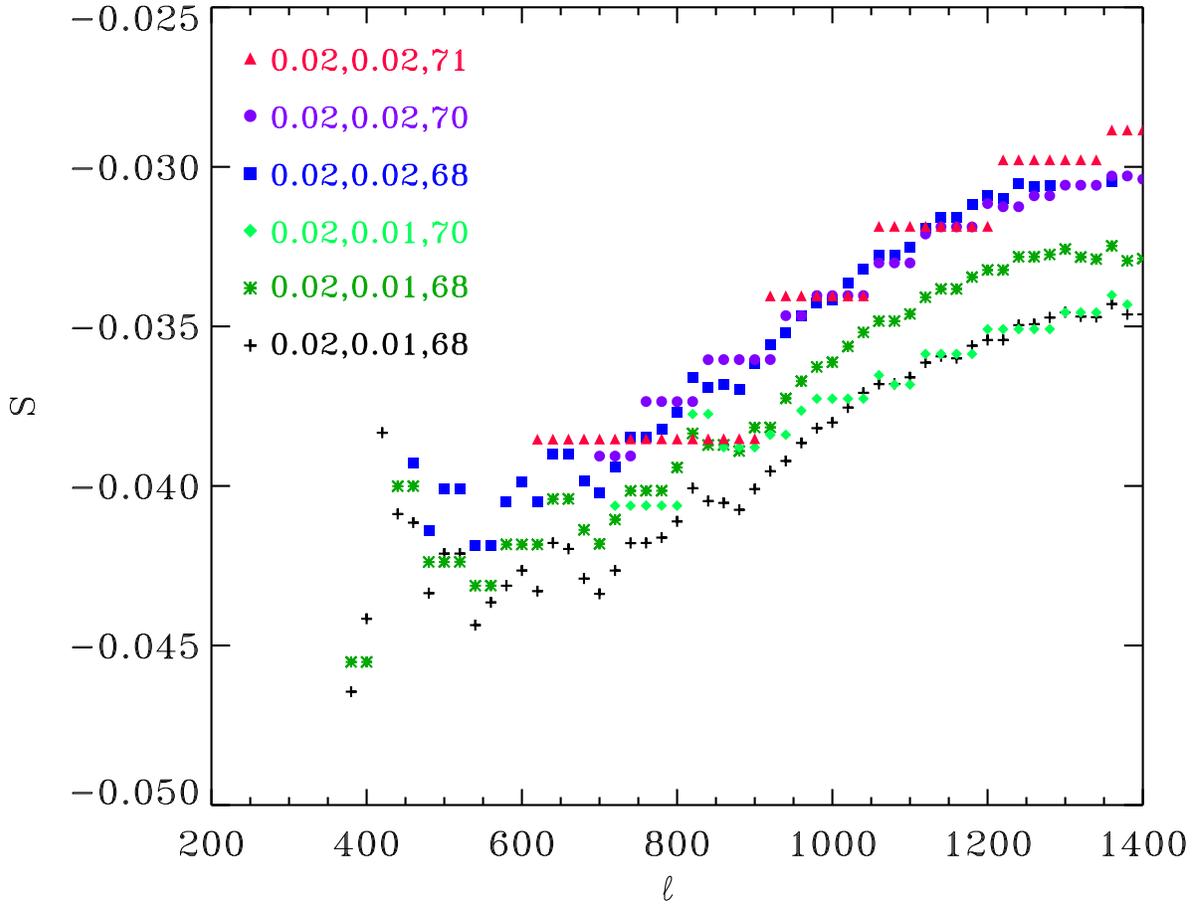}
\caption{The variation in the asymmetry parameter as a function of the flow velocity and background
and is obtained from the fitting of azimuthally averaged synthetic spectra. 
The three values mentioned against each of the symbols represent $U_x$, $U_y$ and $B_1$ component, respectively.
A value of 0.01 for $U_x$ and $U_y$ corresponds to actual velocity of 43.72 m/s.  
\label{fig5}} 
\end{figure}

Although it is not easy to understand these differences due to the 
different functional form of the model profiles to fit different spectra 
(six parameters are fitted in azimuthal spectra as opposed to 13 in the 
three-dimensional spectra), we have made a preliminary attempt by constructing synthetic 3D spectra
with $S = -0.05$. This spectra are then azimuthally averaged and fitted
to calculate the asymmetry in peak profiles. Because of non-zero values of
velocities ($U_x, U_y$) the process of azimuthal averaging also introduces
some asymmetry and hence the value of $S$ in azimuthally averaged spectra
is not the same as that used in preparing the 3D spectra.
The effect of varying the  flow velocity ($U_x$ \& $U_y$) and background term, 
$B_1$ is shown in Fig~\ref{fig5}. 
The synthetic spectra were constructed with a uniform amplitude $A_0=50$,
width $w_0=40\;\mu$Hz and included only the f-mode ridge. 
The background in synthetic spectra included only the first term with
different values of $B_1$. The effect of velocity increases with $k$ and
hence the shift in asymmetry from input value also increases with $k$.
Further, since $A_0$ and $B_1$ are constant, the peak to background
ratio increases with $k$.
It is clear that even when the flow velocities are increased beyond their
normal values, the maximum change in $S$ is about
than 0.02 at the highest value of $\ell$ and rather large velocity is required to make significant 
differences. Similar results are obtained when spectra with different values
of $S$ were used. In all cases azimuthal averaging tends to increase the value
of $S$, though the effect may be small.

\subsection{Source of the asymmetry}
Peak asymmetry is believed to have a contribution from the acoustic source location and a contribution
from the effects of correlated background noise from the source. 
\cite{rv97} have argued that the line asymmetries occur due to the modes correlated with velocity background
while to reach the same effect, \cite{nigamall98} have proposed  a small amount of intensity-correlated 
background.   The amount of coherent correlated background, 
as well as the need of this component to model both $V$ and \ic, has also been debated (e.g. \cite{straus01}).
  
\begin{figure}           
\centering
\plotone{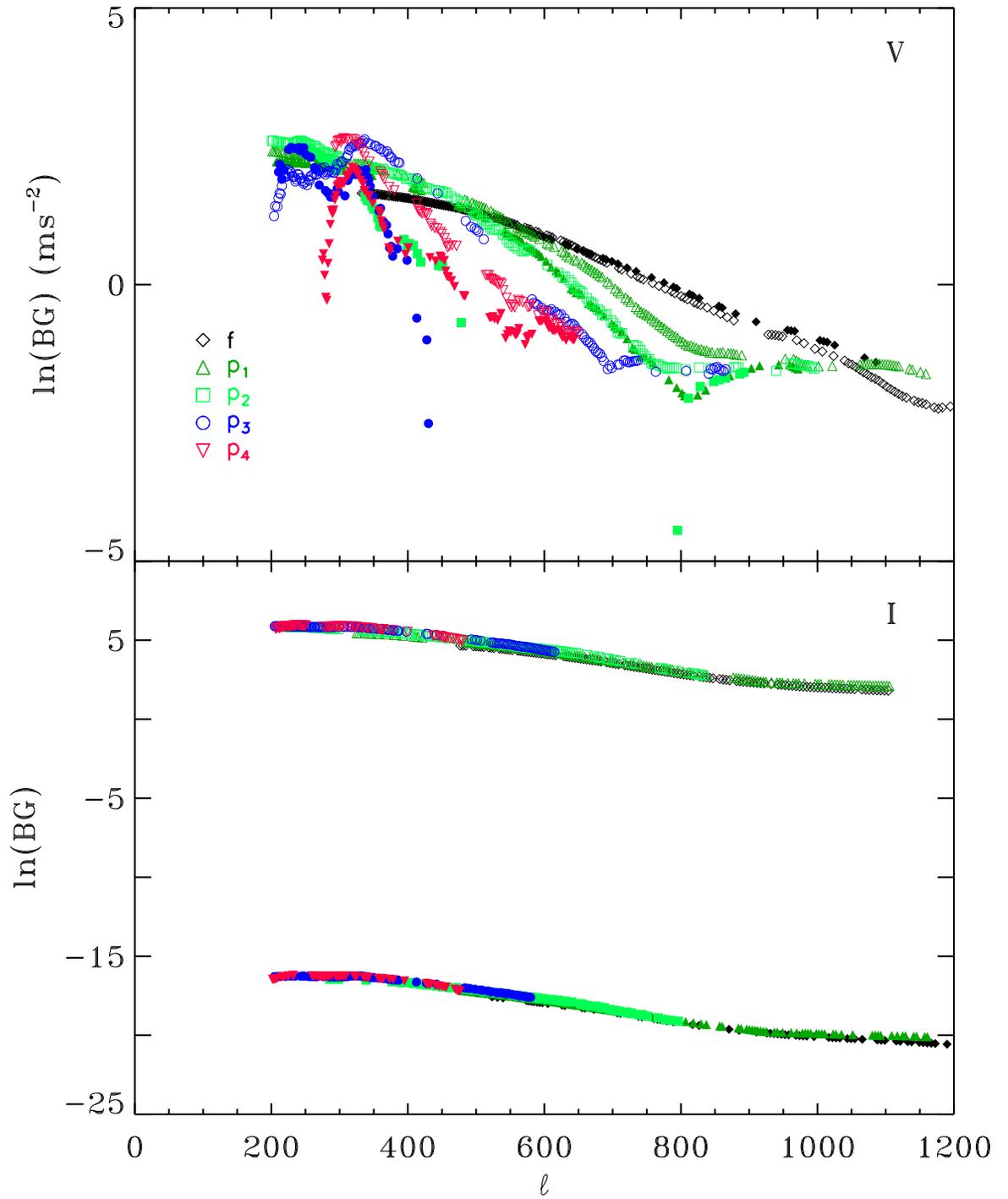}
\caption{Comparison of the background power  obtained from symmetric (filled symbols) 
and asymmetric (open symbols) fits  to the disk center spectra. 
The upper panel 
is for velocity and bottom panel is for continuum intensity. 
 \label{fig6}} 
\end{figure}

In order to understand the role of the background power in the reversal of the 
asymmetry between $V$ and \ic, we plot the background power as a function of degree  $\ell$ of the mode 
in Figure~\ref{fig6}. In the same plot, we also show the background power obtained from the symmetric fits
(open symbols). It is clear that the background power obtained from symmetric and asymmetric 
fitting of the velocity spectra 
(upper panel) closely agree with each other. In contrast the intensity spectrum returns a 
significantly higher background power when fitted with the asymmetric line profile 
compared to the symmetric profile. This indicates that the line asymmetry is reversed in the intensity. 
Since granulation contrast decreases from the disk center 
to the limb, we also expect the background power to decrease near the limb. But no  
 significant variation in power is noticed from the disk center to the limb which supports 
the idea that only a part of the background is correlated with the intensity oscillations \citep{nigam99, bh04}.       

\subsection{Frequency Differences}
\begin{figure} 
\plotone{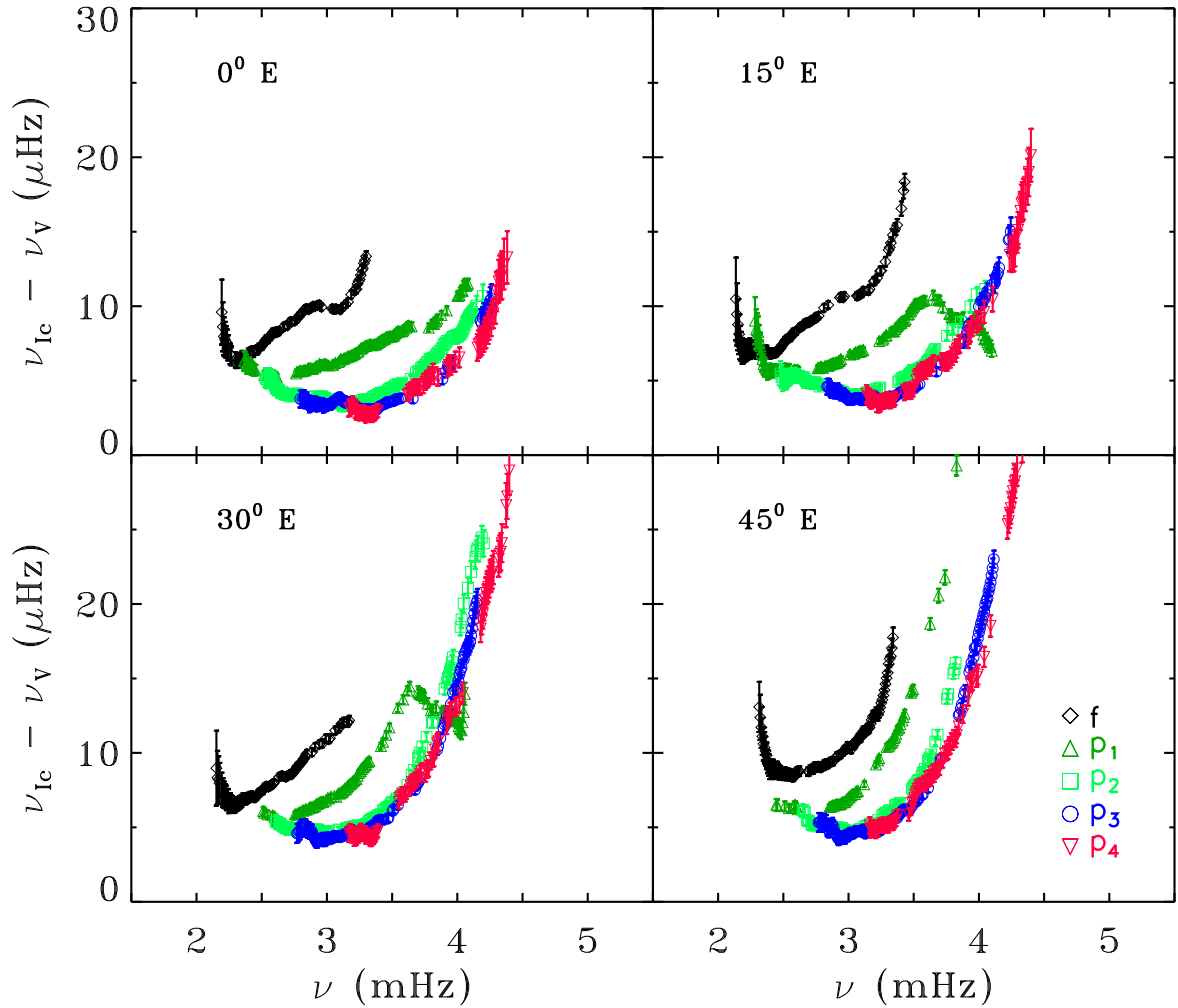}
\caption{Frequency shifts between velocity and continuum intensity modes fitted using 
symmetric profile at four different 
longitudes. The locations are marked in each panel. 
\label{fig7}}
\end{figure}

The frequency differences from the symmetric fits to both $V$ and \ic spectra at four 
different longitudes are plotted in Figure~\ref{fig7}. The frequencies obtained from the 
intensity spectrum are systematically higher than the frequencies from the velocity spectrum,
as one would expect due to the reversal of the line asymmetry. 
For the disk center spectra, the differences, in general,
 are on the order of 10~$\mu$Hz or less up to $\nu \le 4$~mHz; the minimum 
difference of about 2--3~$\mu$Hz is seen for $p_2$ and higher radial order modes in the 5--min band.  
Since near the limb we primarily  observe the horizontal velocity field which is dominated by granulation, 
it is expected that the frequency difference between intensity and velocity should decrease near the limb
due to a decrease in correlation with oscillations. 
But we find that the difference increases systematically from the disk  center to the limb. 
 This again supports the idea that only a part of the background is 
correlated with the intensity oscillations. 

\begin{figure} 
\plotone{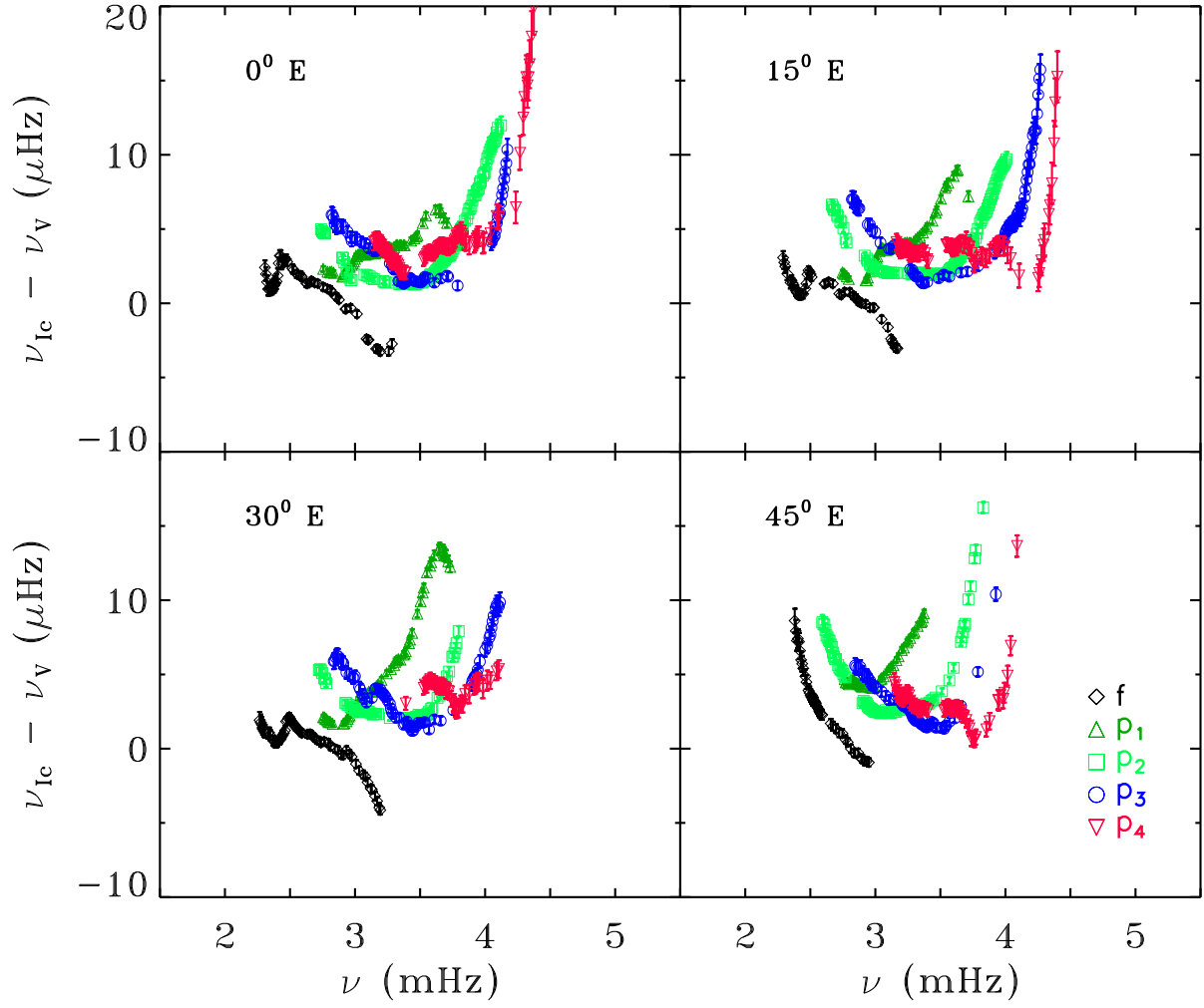}
\caption{Same as Figure~\ref{fig7} but for modes obtained from the fit 
using asymmetric profiles.
\label{fig8}}
\end{figure}

The use of asymmetric profiles to fit the power spectra results in a better determination 
of the mode frequencies and as a result the  frequency differences 
between velocity and intensity is significantly diminished (Figure~\ref{fig8}).  
Similar to fits with symmetric profiles, the differences are mostly positive 
as the frequencies obtained from intensity fluctuations 
are higher than those from the velocity. However, the {\it f}-modes show some exception at the high end of 
the frequency where the peak frequencies observed in velocity are marginally higher  
(and clearly larger than the estimated errors) than those observed in intensity. 
The differences between higher order radial modes, mainly $p_3$ and $p_4$, are slightly
bigger but these modes also have a large uncertainty in the fitted 
frequencies as seen from the error estimates. Although subtle variations are evident between different longitudes, 
the frequency differences do not reveal any systematic trend between 
the disk center and 45\degr \ longitude.

\begin{figure}           
\centering
\plotone{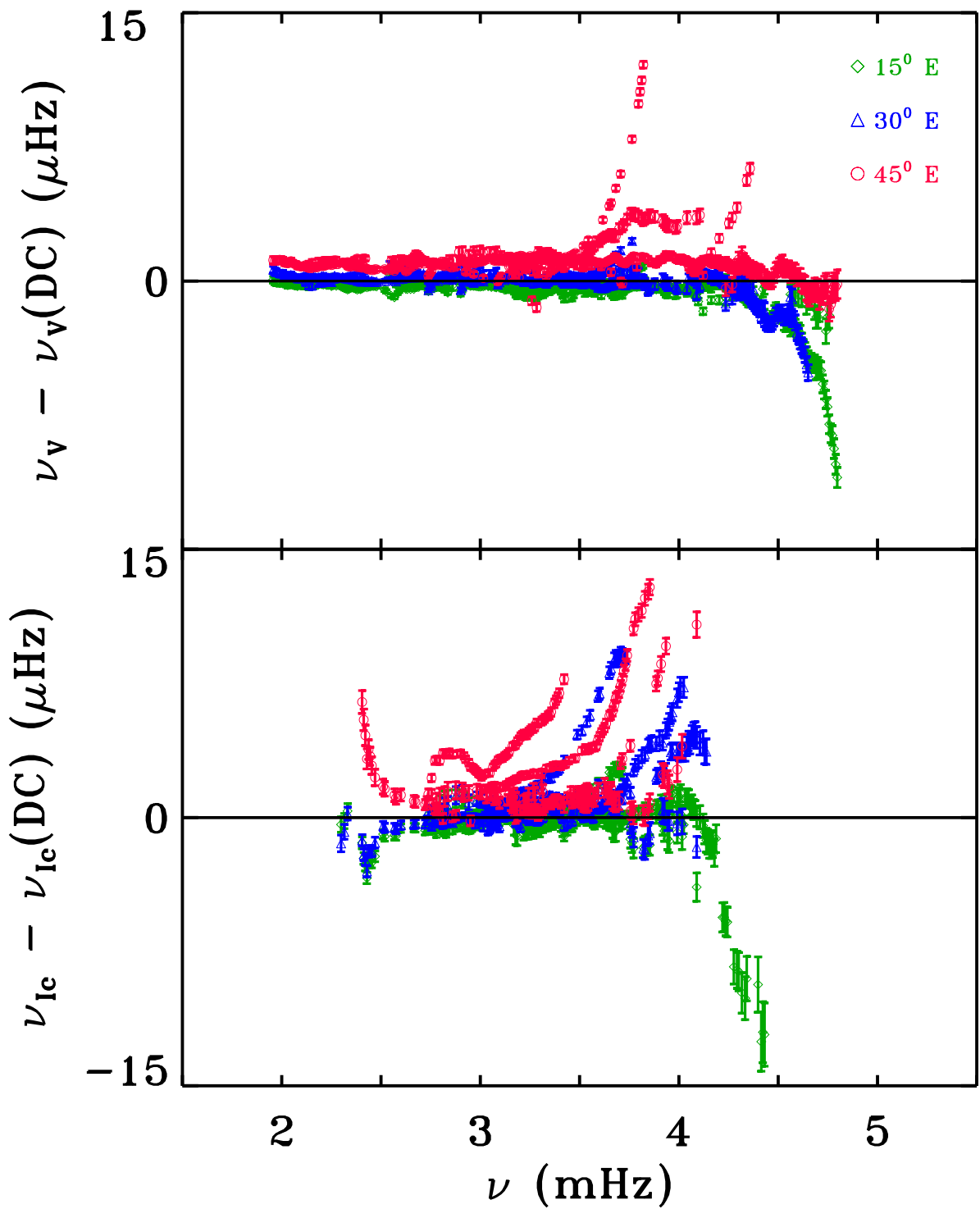}
\caption{Frequency shifts between velocity (top panel) and continuum intensity (bottom panel) 
modes at three different longitudes  with respect to disk center obtained from the fit 
using asymmetric profiles. The locations 
are marked in the top panel. 
 \label{fig9}} 
\end{figure}

Although the asymmetric fitting formula reduces the frequency differences between the velocity 
and intensity spectra, there still remains some systematic differences of the order of 
5\todash10~$\mu$Hz,  which is much higher than the uncertainty in estimated errors. 
 \cite{bab01} have  reported a similar result based on the analysis of 
azimuthally averaged spectra and have argued that 
the assumed profiles does not adequately describe the velocity spectra where the signal-to-noise 
ratio is high.   However, our result  
indicates that the form of asymmetry used for intensity spectra  
 may not be a correct representation of the real profiles. To demonstrate 
this, we show the difference in eigenfrequencies between identical modes obtained from the power 
spectra at different longitudes with respect to the spectra at the disk center in Figure~\ref{fig9}. 
The frequency difference in $V$ shows little variation in longitude for 
frequencies less than 3.5~mHz. Variations between intensity spectra are higher than seen in 
velocity. Moreover, the shapes of the curve in  Figure~\ref{fig9} for velocity are similar
as are the shapes of the curves in intensity, leading to the frequency differences shown in
Figure~\ref{fig8}. Therefore it appears that the use of an asymmetric line profile provides 
more accurate estimates of the eigenfrequencies of solar oscillation for velocity rather than for 
intensity. In addition,  fitting the intensity spectra is more challenging than  fitting the 
velocity spectra due to low amplitudes and low signal-to-noise ratio. 
The number of modes fitted from the intensity spectra clearly supports this conjecture. 

\section{Conclusions}
Using the local helioseismic technique of ring diagram analysis we have studied the high-degree 
{\it f}- and {\it p}-mode frequency differences between velocity and continuum 
intensity data obtained from the MDI instrument during a quiet period.   
Since the spectra are known to be asymmetric, we  fitted 
the three-dimensional spectra with an asymmetric model profile 
based on the known form of the asymmetry observed in velocity and 
intensity. The asymmetry obtained from fitting the velocity spectra agrees with the 
previous results while that of intensity shows a frequency dependent behavior. 
We further notice significant differences between the asymmetry obtained from fitting the 
three-dimensional and azimuthally averaged power spectra.

Fitting the three-dimensional disk center power spectra with a symmetric Lorentzian profiles leads to
frequency differences on the order of 10~$\mu$Hz or less up to $\nu \le$ 4~mHz  between intensity 
and velocity. 
This difference is found to be increasing systematically from the disk center to the limb. 
 The use of asymmetric profiles leads to frequency differences that are smaller 
than the differences resulting from the symmetric fits. 
However, systematic differences still remain at the high end of the frequency range, 
mostly above 4~mHz. We demonstrate that this difference arises from the fitting of the 
intensity rather than the velocity spectra. We speculate that 
the form of asymmetry used in the model for the intensity spectra is over-simplified and does not 
adequately represent the real profile at all frequency ranges. 

We also conclude 
that the frequency differences between velocity and intensity do not vary significantly 
from the disk center to the limb. However, variations between 
intensity spectra are higher than seen in velocity. This supports the idea that only a part of the background is 
correlated with the intensity oscillations. In this context several authors (e.g.  \cite{severino01, bhc04, toutain06}) 
 have advocated the importance of fitting the intensity-velocity cross spectrum along with a 
 multicomponent model of the coherent background. This provides an impetus for future work using the 
high-resolution observations from {\it Solar Dynamics Observatory}.

\acknowledgments
This study uses data from MDI/SOHO. 
SOHO is a mission of international cooperation between
ESA and NASA.  This work was supported by NASA grant NNG 5-11703 and NNG 05HAL41I.

\clearpage

\end{document}